\documentclass[a4paper,11pt]{article}
\pdfoutput=1 

\usepackage{jheppub} 
\usepackage{bbm,bm,graphicx,mathtools,color,slashed,hyperref}
\usepackage{amssymb,amsmath}

\graphicspath{{./Figures/}}

\newcommand{\diff}{\mathrm{d}}

\newcommand{\ve}{\varepsilon}
\newcommand{\Diff}{{\mathcal{D}}}

\newcommand{\tr}{\mathrm{tr}}
\newcommand{\im}{\mathrm{i}}

\newcommand{\calZ}{\mathcal{Z}}

\newcommand{\rmd}{\mathrm{d}}
\newcommand{\rme}{\mathrm{e}}

\newcommand{\rmA}{\mathrm{A}}
\newcommand{\rmB}{\mathrm{B}}
\newcommand{\rmL}{\mathrm{L}}
\newcommand{\rmR}{\mathrm{R}}
\newcommand{\rmV}{\mathrm{V}}
\newcommand{\rep}{\mathsf{rep}}
\newcommand{\fd}{\mathsf{fd}}
\newcommand{\adj}{\mathsf{adj}}
\newcommand{\sym}{\mathsf{sym}}
\newcommand{\asym}{\mathsf{asym}}

\preprint{YITP-22-45}

\title{Semiclassics with 't~Hooft flux background for QCD with $2$-index quarks}

\author[1]{Yuya Tanizaki,}

\affiliation[1]{Yukawa Institute for Theoretical Physics,  Kyoto University, Kyoto 606-8502, Japan}

\emailAdd{yuya.tanizaki@yukawa.kyoto-u.ac.jp}

\author[2]{Mithat \"{U}nsal}

\affiliation[2]{Department of Physics, North Carolina State University, Raleigh, NC 27607, USA}

\emailAdd{unsal.mithat@gmail.com}

\abstract{
We study quantum chromodynamics including the two-index symmetric or anti-symmetric quark (QCD(Sym/ASym)) on small $\mathbb{R}^2\times T^2$ with a suitable magnetic flux. 
We first discuss the 't~Hooft anomaly of these theories and claim that discrete chiral symmetry should be spontaneously broken completely to satisfy the anomaly matching condition. 
The $T^2$ compactification with the magnetic flux preserves the 't~Hooft anomaly, and the $2$d effective theory is constrained by the same anomaly of $4$d QCD(Sym/ASym). 
We demonstrate the spontaneous breakdown of chiral symmetry using the dilute gas of center vortices, which confirms the prediction of the 't~Hooft anomaly. 
We also find that each vacuum maintains the charge conjugation symmetry, and this gives affirmative support for the nonperturbative large-$N$ orientifold equivalence between QCD(Sym/ASym) and $\mathcal{N}=1$ supersymmetric $SU(N)$ Yang-Mills theory.
}

\begin{document}
\maketitle


\section{Introduction}\label{sec:introduction}

In our recent work~\cite{Tanizaki:2022ngt}, we have proposed a new semiclassical formalism to study four-dimensional non-Abelian gauge theories by compactifying the theory on small $\mathbb{R}^2\times T^2$ including 't~Hooft flux. 
Our framework provides the weak-coupling description of confinement and reproduces many qualitative features of strongly coupled confining vacua, and thus it is reasonable to expect that its dynamics show adiabatic continuity to the strong dynamics on $\mathbb{R}^4$. 
  
Technically, the 't Hooft flux lifts the classical moduli space of gauge holonomies on $T^2$ completely~\cite{tHooft:1979rtg}, and the $0$-form center symmetry is unbroken. 
As the $T^2$ size is much smaller than the strong scale,  the center-symmetric vacuum is stable under perturbative corrections.
On the other hand, the $1$-form center symmetry is spontaneously broken at the classical level,  and we need to consider the nonperturbative corrections to restore it.

The center vortex~\cite{tHooft:1977nqb, Cornwall:1979hz, Nielsen:1979xu, Ambjorn:1980ms} is the key player that dominates the leading nonperturbative correction. 
In this $T^2$ compactified setup with 't~Hooft flux, center vortices are realized as the BPS configuration having the fractional topological charge $Q = \pm  \frac{1}{N}$, and its classical action is given by $S= \frac{S_I} {N}$, where $S_I= \frac{8 \pi^2}{g^2}$  is the 4d instanton action~\cite{Gonzalez-Arroyo:1998hjb, Montero:1999by, Montero:2000pb}. 
The distinguishing feature of the center vortex is its nontrivial mutual statistics with Wilson loops. 
We can show that the dilute-gas approximation of center vortices describes the confinement for $1$-form center symmetry, and this provides the first semiclassical computations of $4$d confining gauge theories based on center vortices.

In this paper, we generalize our earlier work to quantum chromodynamics (QCD) with two-index representation fermions, e.g. symmetric or anti-symmetric representation, and we denote them as QCD(Sym/ASym). 
QCD(ASym) provides an alternative large-$N$ limit of QCD because of the simple kinematic fact that fundamental and anti-symmetric representations are the same for $N=3$. 
Unlike the 't~Hooft large-$N$ limit~\cite{tHooft:1973alw}, quark loops are not suppressed compared with gluon loops in QCD(ASym), and this feature is similar to the large-$N$ limit of QCD with $N_f$ fundamental quarks with $N_f/N$ fixed. 
QCD(Sym/ASym) received much attention in the context of orientifold equivalence with ${\cal N}=1$ supersymmetric Yang-Mills theory~\cite{Armoni:2003gp, Armoni:2004ub, Unsal:2006pj}.  

We examine the nonperturbative dynamics of   $N_f=1$  QCD(Sym/ASym) by using compactification with magnetic flux. 
We shall demonstrate that the discrete chiral symmetry is spontaneously broken completely, by performing the dilute gas approximation with center vortices dressed with gapless fermions. 
By adding a soft mass deformation, we obtain the multi-branch structure of the ground states as in the case of confining vacua of pure Yang-Mills theory. 
A remarkable feature here is that even though the theta angle is fractionalized as $\theta/N$, the dilute gas of center vortices produces the correct $N+2$ (and $N-2$) vacua and multi-branched structure once a soft mass term is added. 
As these theories are non-supersymmetric, there have not been powerful techniques to study QCD(Sym/ASym) directly (except with  double-trace deformations  on $\mathbb R^3 \times S^1$~\cite{Shifman:2008ja}), and thus our result provides a novel insight into their dynamics and the validity of nonperturbative orientifold equivalence.


\section{QCD with \texorpdfstring{$2$}{2}-index (anti-)symmetric representation}

In this section, we discuss the dynamics of $1$-flavor QCD(Sym) and QCD(ASym). 
Since the analyses on these theories become almost parallel, we pay attention to QCD(Sym) first, and then we briefly discuss the case of QCD(ASym) later. 

\subsection{Global symmetry and 't~Hooft anomaly of QCD(Sym)}
\label{sec:anomaly_QCD(S)}

We consider the $4$d $SU(N)$ gauge theory coupled to the Dirac fermion with $2$-index symmetric representation. 
This model is denoted as QCD(Sym), and we set $N\ge 3$ in the following discussion.\footnote{For $N=2$, the $2$-index symmetric representation is nothing but the adjoint representation, which is a real representation. For $N\ge 3$, the representation is a complex representation. This causes the significant difference on the global symmetry, so we only consider the case with $N\ge 3$. } 
The Lagrangian of QCD(Sym) is given by 
\begin{equation}
S=\frac{1}{g^2}\int \tr(f\wedge \star f)+\frac{\im \theta}{8\pi^2}\int \tr(f\wedge f)+\int \overline{\psi}\gamma_\mu D_\mu \psi,    
\end{equation}
where $f=\diff a+\im a\wedge a$ is the $SU(N)$ gauge field strength, and $D=\diff +\im a$ is the covariant derivative. 
Since $\psi$ belongs to the $2$-index symmetric representation, the fermion kinetic term takes the following form:
\begin{equation}
    \overline{\psi}\gamma_\mu D_\mu \psi =\overline{\psi}\gamma_\mu\left(\partial_\mu+\sym(\im a_\mu)\right)\psi. 
\end{equation}

The classical theory has the chiral symmetry, $U(1)_\rmL\times U(1)_\rmR\simeq [U(1)_\rmV\times U(1)_\rmA]/\mathbb{Z}_2$, but its axial part is explicitly broken by the Adler-Bell-Jackiw (ABJ) anomaly.  
Under the $U(1)_\rmA$ transformation, 
\begin{equation}
    \psi\mapsto \rme^{\im \alpha \gamma_5}\psi, \, 
    \overline{\psi}\mapsto \overline{\psi}\,\rme^{\im \alpha \gamma_5}, 
\end{equation}
the fermionic path-integral measure is changed as 
\begin{equation}
    \Diff \overline{\psi}\Diff \psi \mapsto 
    \Diff \overline{\psi} \Diff \psi 
    \exp\left(\im \frac{2\alpha}{8\pi^2}\int \tr_{\sym}(f\wedge f)\right),
\end{equation}
where $\tr_{\sym}$ denotes the trace operation over the $2$-index symmetric representation. 
Using the identity, $\tr_{\sym}(T^a T^b)=(N+2)\tr(T^a T^b)$, we find that 
\begin{equation}
    \Diff \overline{\psi}\Diff \psi \mapsto 
    \Diff \overline{\psi} \Diff \psi 
    \exp\left(\im \frac{2(N+2)\alpha}{8\pi^2}\int \tr(f\wedge f)\right), 
\end{equation}
and then the measure is invariant if $\alpha\in \frac{2\pi}{2(N+2)}\mathbb{Z}$. 
Therefore, the axial symmetry $U(1)_\rmA$ is explicitly broken to its discrete subgroup, $\bigl(\mathbb{Z}_{2(N+2)}\bigr)_\rmA$. 

We also need to care about the periodicity of the vector symmetry and its relation to the $1$-form symmetry. 
When the matter fields are absent, the model becomes the $SU(N)$ pure Yang-Mills theory, which enjoys the $\mathbb{Z}_N^{[1]}$ symmetry, or the center symmetry. 
When the $2$-index fermion is introduced, the vector-like symmetry including the $1$-form symmetry is given by 
\begin{equation}
    G_\rmV=\left\{ \begin{array}{cc}
    U(1)/\mathbb{Z}_{\frac{N}{2}}\times \mathbb{Z}_2^{[1]} &  (\mathrm{even}\, N),\\
    U(1)/\mathbb{Z}_N & (\mathrm{odd}\, N).  
    \end{array}\right.
    \label{eq:vector_sym}
\end{equation}
For even $N$, $\mathbb{Z}_2^{[1]}\subset \mathbb{Z}_{N}^{[1]}$ survives as the global symmetry since the $2$-index representation is neutral, while for odd $N$, the $1$-form symmetry is completely gone. 
This also indicates that the part of the $U(1)_\rmV$ transformation has an overlap with a part of the center subgroup of the $SU(N)$ gauge transformations, and the faithful $U(1)$ symmetry group has the fractional periodicity in terms of the quark language. 

In order to discuss the 't~Hooft anomaly, let us introduce the background gauge field for the vector-like symmetry $G_{\rmV}$ and we shall check if the discrete chiral symmetry becomes anomalous (For  recent developments of 't~Hooft anomalies, see Refs.~\cite{Gaiotto:2017yup, Tanizaki:2017bam, Komargodski:2017dmc, Komargodski:2017smk, Shimizu:2017asf, Gaiotto:2017tne}). 
Taking the above expression for the symmetry structure naively, the structure of the background gauge fields seems to depend on $N \bmod 2$. 
However, thanks to the subtle $2$-group extension for even $N$, the background gauge fields become almost identical to the case with those for odd $N$. 
In order to see this, it is convenient to introduce the $U(1)$ background gauge field $A$ first, and then the fermion kinetic term becomes
\begin{equation}
    \overline{\psi}\gamma^\mu(\partial_\mu+\im (\sym(a_\mu)+ A_\mu))\psi. 
\end{equation}
This has the $\mathbb{Z}_N^{[1]}$ symmetry as it is invariant under the following transformation, 
\begin{equation}
    a\mapsto a+\frac{1}{N}\ve,\quad A\mapsto A-2\frac{1}{N}\ve, 
\end{equation}
where $\frac{1}{N}\ve$ is the flat $\mathbb{Z}_N$ gauge field. 
Therefore, we can introduce the $\mathbb{Z}_N$ $2$-form gauge field, which is realized as the $U(1)$ $2$-form gauge field with the constraint, 
\begin{equation}
    NB=\diff C, 
\end{equation}
where $C$ is the auxiliary $U(1)$ gauge field. 
To have the coupling between $(B,C)$ and QCD(Sym), we use the trick that $a$ is promoted to the $U(N)$ gauge field $\widetilde{a}$ given by~\cite{Kapustin:2014gua} 
\begin{equation}
    \widetilde{a}=a+\frac{1}{N}C\bm{1}_N.
\end{equation}
This should be understood as the local expression, and what it actually means is that the path integral $\int \Diff \widetilde{a}$ is performed over the $U(N)$ gauge fields with the constraint 
\begin{equation}
    \tr(\widetilde{a})=C. 
\end{equation}

We postulate the $U(1)$ $1$-form gauge transformation, 
\begin{equation}
    B\mapsto B+\diff \lambda, \, C\mapsto C+N\lambda, 
\end{equation}
and thus 
\begin{equation}
    \widetilde{a}\mapsto \widetilde{a}+\lambda \bm{1}_N,\, 
    A\mapsto A-2\lambda. 
\end{equation}
All of these operations are valid for both even and odd $N$, and the background gauge fields for $G_\rmV$  are successfully coupled to QCD(Sym). 

How can we see the difference of the symmetry group in the language of these background gauge fields? 
To see this, we have to note that the gauge-invariant $U(1)$ field strength is given by 
\begin{equation}
    \diff A+2 B, 
\end{equation}
since $A$ is transformed under the $U(1)$ $1$-form gauge transformation. 
Since $B$ is quantized in $\frac{1}{N}$, this does not satisfy the proper Dirac quantization for $U(1)$ gauge fields. 
If we denote the corresponding $U(1)$ gauge field with the proper Dirac quantization as $A_\rmB$, then its field strength should be given by~\cite{Tanizaki:2018wtg}
\begin{equation}
    \diff A_\rmB=\left\{\begin{array}{cc}
    \frac{N}{2}(\diff A+2B) & \quad (\mbox{even}\, N), \\
    N(\diff A+2B) & \quad (\mbox{odd}\, N), 
    \end{array}\right.
\end{equation}
and we here encounter the distinction between even and odd $N$. 
The coefficient in front on the right-hand-side discriminates that the minimal gauge-invariant baryonic operator is given by $\psi^{\otimes (N/2)}$ for even $N$ while $\psi^{\otimes N}$ for odd $N$. 
When $N$ is a multiple of $4$, the baryon operator is bosonic, which corresponds to the fact that the fermion parity $(-1)^F$ is a part of $SU(N)$ gauge redundancy for $N\in 4\mathbb{Z}$.

Next, we discuss the 't~Hooft anomaly for QCD(Sym) at the massless point. 
Under the presence of the $G_\rmV$ gauge field, the index of the Dirac operator $\slashed{D}=\gamma^\mu(\partial_\mu+\im(\sym(\tilde{a}_\mu)+A_\mu))$ is given by 
\begin{align}
    \mathrm{Ind}(\slashed{D})&=\frac{1}{8\pi^2}\int\left(\tr_{\sym}((\widetilde{f}-B)^2)+\dim(\sym)(\diff A+2B)^2\right)\nonumber\\
    &=\frac{1}{8\pi^2}\int\left((N+2)\tr((\widetilde{f}-B)^2)+\frac{N(N+1)}{2}(\diff A+2B)^2\right) \nonumber\\
    &=\frac{N+2}{8\pi^2}\int \tr(\widetilde{f}\wedge \widetilde{f})+\frac{1}{8\pi^2}\int\left(\frac{N(N+1)}{2}(\diff A)^2+2(N+1)\diff A\wedge N B+(NB)^2\right). 
\end{align}
Under the discrete axial transformation, 
\begin{equation}
    \bigl(\mathbb{Z}_{2(N+2)}\bigr)_\rmA:\psi\mapsto \rme^{\frac{2\pi }{2(N+2)}\im\gamma_5}\psi,\, 
    \overline{\psi}\mapsto \overline{\psi}\,\rme^{\frac{2\pi }{2(N+2)}\im\gamma_5}, 
\end{equation}
the partition function with the background gauge fields, $\calZ[A,B]$, acquires the overall phase factor,
\begin{align}
    \bigl(\mathbb{Z}_{2(N+2)}\bigr)_\rmA: &\calZ[A,B]\mapsto \exp\left(\frac{2\pi\im}{(N+2)}\mathrm{Ind}(\slashed{D})\right)\calZ[A,B]\nonumber\\
    &\quad =\rme^{\frac{2\pi \im}{8\pi^2(N+2)} \int\left(\frac{N(N+1)}{2}(\diff A)^2+2(N+1)\diff A\wedge N B+(NB)^2\right)}\calZ[A,B]. 
    \label{eq:qcdsym_anomaly}
\end{align}
This is a special case of the baryon-color(-flavor) anomaly discussed in Refs.~\cite{Anber:2019nze, Anber:2021lzb}. 
If we assume that the low-lying states are described by color-singlet hadrons and that the anomaly is matched by spontaneous breaking of the axial symmetry, we can conclude the complete breakdown,
\begin{equation}
    \bigl(\mathbb{Z}_{2(N+2)}\bigr)_\rmA\xrightarrow{\mathrm{SSB}} \mathbb{Z}_2, 
\end{equation}
and there have to be $(N+2)$ distinct vacua.

\subsection{Semiclassical study of QCD(Sym) on small \texorpdfstring{$\mathbb{R}^2\times T^2$}{R2xT2} with magnetic flux}

Having the knowledge on the 't~Hooft anomaly of QCD(Sym), we construct its anomaly-preserving $T^2$ compactification and study its $2$d effective field theory.
This can be achieved by introducing the magnetic flux along small $T^2$ following Ref.~\cite{Tanizaki:2022ngt}, and it allows us to study the dynamics of QCD(Sym) in a controllable semiclassical calculation. 
Therefore, we take the $4$d spacetime to be $M\times T^2$, where $M$ is a sufficiently large closed manifold and the size of $T^2$ is set to be much smaller than the strong scale. 
We shall see that the semiclassical computation predicts the complete spontaneous breakdown of the discrete chiral symmetry. 

\subsubsection{Tree-level analysis}

We would like to keep the 't~Hooft anomaly as intact as possible under the $T^2$ compactification~\cite{Tanizaki:2017qhf, Tanizaki:2017mtm, Yamazaki:2017dra}. 
One of the simplest possibilities for this purpose is to set
\begin{equation}
    A=0,\,\, B=\frac{2\pi}{N} \frac{\diff x_3\wedge \diff x_4}{L^2}, 
    \label{eq:magnetic_flux}
\end{equation}
where $x_3\sim x_3+L$ and $x_4\sim x_4+L$. 

To make the connection with the conventional approach, let us write down the Dirac equation and the boundary condition on $T^2$ explicitly. 
The boundary condition is denoted as 
\begin{align}
    \psi(x_3+L,x_4)&=\sym(g_3(x_4)^\dagger)\rme^{-\im \phi_3(x_4)}\psi(x_3,x_4), \\
    \psi(x_3,x_4+L)&= \sym(g_4(x_3)^\dagger) \rme^{-\im \phi_4(x_3)} \psi(x_3,x_4). 
\end{align}
Since we have introduced the minimal 't~Hooft flux, $\int_{T^2}B=\frac{2\pi}{N}$, the color transition functions must satisfy, 
\begin{equation}
    g_3^\dagger(L)g_4^\dagger(0)=g_4^\dagger(L)g_3^\dagger(0)\rme^{\frac{2\pi}{N}\im}.
\end{equation}
Then, the uniqueness of the matter wavefunction requires that the $U(1)$ transition functions satisfy 
\begin{equation}
    \exp\Bigl(\im \bigl\{(\phi_4(L)-\phi_4(0))-(\phi_3(L)-\phi_3(0))\bigr\}\Bigr)=\exp\left(-\frac{4\pi}{N}\im\right). 
\end{equation}

The net $U(1)$ gauge field is given by $\widetilde{A}=A+\frac{2}{N}C$, and we choose the gauge so that
\begin{equation}
    \widetilde{A}=\frac{4\pi}{N}\frac{x_3\diff x_4}{L^2}. 
\end{equation}
The $U(1)$ transition functions are given by $A(x_3+L,x_4)=A(x_3,x_4)+\diff \phi_3(x_4)$ and $A(x_3,x_4+L)=A(x_3,x_4)+\diff \phi_4(x_3)$, and we obtain 
\begin{equation}
    \phi_3(x_4)=\frac{4\pi}{N} \frac{x_4}{L},\, \phi_4(x_3)=0, 
\end{equation}
and the transition functions satisfy the above constraint. 
For the $SU(N)$ color transition functions, we can perform the gauge transformation so that 
\begin{equation}
    g_3(x_4)=S, \, g_4(x_3)=C, 
\end{equation}
where $C_{nm}\propto \rme^{2\pi \im n/N}\delta_{nm}$ and $S_{nm}\propto \delta_{n+1,m}$ are the clock and shift matrices. 
This gives a concrete realization of the background magnetic flux~\eqref{eq:magnetic_flux}. 
With this choice of the transition functions, the $SU(N)$ gauge fields on $T^2$ obey the boundary conditions, $a(x_3+L,x_4)=S^\dagger a(x_3,x_4)S$ and $a(x_3,x_4+L)=C^\dagger a(x_3,x_4)C$, and we can specifically have $a=0$ as the classical minimum. 
Therefore, for the classical vacuum, the gauge holonomies and the transition functions are identified, and we have 
\begin{equation}
    P_3=S, \, P_4=C. 
\end{equation}
Since the $SU(N)$ gauge group is Higgsed to $\mathbb{Z}_N$, there are no perturbatively gapless spectrum in the gauge sector.

Next, we have to discuss the gapless modes of the fermionic fields. 
To solve it, we have to solve the zero-mode Dirac equation,  
\begin{equation}
    \left[\gamma_3\partial_3+\gamma_4 \left(\partial_4+\im \frac{4\pi}{N}\frac{x_3}{L^2}\right)\right]\psi(x_3,x_4)=0,
    \label{eq:Dirac_zero_mode}
\end{equation}
with the above boundary condition. 
We introduce the $2$d chirality operator in $T^2$ as $\im \gamma_3\gamma_4$, which satisfies $(\im \gamma_3\gamma_4)^2=1$. 
To find the number of zero modes, we can use the $2$d index theorem, which gives 
\begin{equation}
    \mathrm{Ind}(D_{T^2})=\dim(\mathrm{Rep})\int_{T^2}\frac{F_{U(1)}}{2\pi}. 
    \label{eq:index_T2}
\end{equation}
In our case, $\dim(\mathrm{Rep})=\frac{N(N+1)}{2}$ and $\int \frac{F}{2\pi}=\frac{2}{N}$, so we have $N+1$ flavors of $2$d Dirac fermions.\footnote{For the fundamental representation, we have $\dim(\mathrm{Rep})=N$ and $\int \frac{F}{2\pi}=\frac{1}{N}$, and we get a single $2$d Dirac fermion. This reproduces the result of our previous analysis on QCD with fundamental quarks~\cite{Tanizaki:2022ngt}. }
We confirm this result in Appendix~\ref{sec:Dirac_zero} by solving the Dirac equation~\eqref{eq:Dirac_zero_mode}.

\subsubsection{Perturbative analysis}

We have $(N+1)$ flavors of $2$d massless fermions, so we naively have $U(N+1)_\rmL\times U(N+1)_\rmR$ chiral symmetry. 
However, this is absent in $4$d, and thus most part of it should be broken by higher-dimensional operators in the $2$d effective theory. 
In $2$d, four-fermion operators are perturbatively marginal, and they can be marginally relevant. 
Therefore, except $U(1)_\rmA$, the axial symmetries would be explicitly broken by integrating out massive Kaluza-Klein (KK) modes.

Let us perform Abelian bosonizations, then we get $N+1$ compact bosons, $\phi_1, \ldots, \phi_{N+1}$. 
The correspondence is given by
\begin{equation}
    \overline{\psi_\rmL^i}{\psi}_\rmR^i\sim \rme^{\im \phi_i},
\end{equation}
and the tree-level effective action becomes 
\begin{equation}
    S_{\mathrm{tree}}=\sum_{i=1}^{N+1}\int_{M}\frac{1}{8\pi}|\diff \phi_i|^2. 
\end{equation}
Since the $U(1)_\rmA$ transformation corresponds to $\phi_i\mapsto \phi_i+2\alpha$, we can introduce the following perturbation without breaking $U(1)_\rmA$:
\begin{equation}
    \Delta S=\int_M\diff^2 x \sum_{ij} J_{ij}\cos(\phi_i-\phi_j).
    \label{eq:4fermi_bosonized}
\end{equation}
This corresponds to the four-fermion vertices in the fermionic theory. 

We note that the one-gluon exchange diagram
\begin{align}
	\includegraphics[scale=0.4]{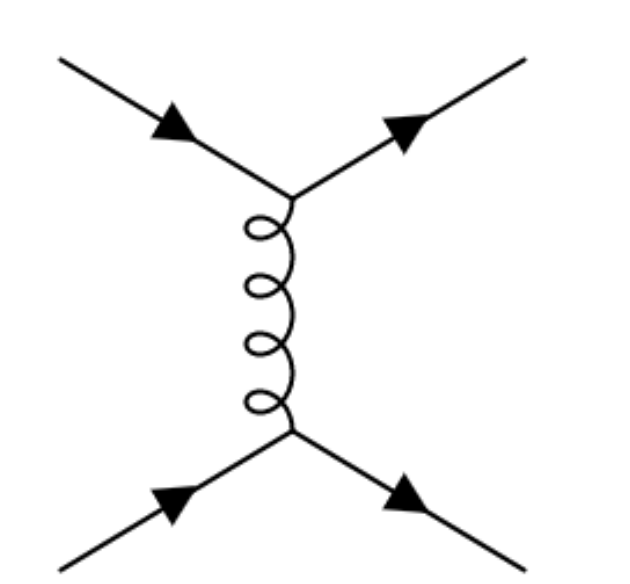}
\end{align}
cannot produce $\Delta S$, and they are generated by the loop diagrams. 
The tree-level $4$d gluon exchange can be decomposed into the $2$d vector-boson exchange and the scalar exchange. 
The exchange of $2$d vector boson gives the vector-vector four-fermion coupling, which modifies the kinetic term in the bosonic description instead of generating the potential term. 
Therefore, we need to consider the exchange of scalar $P_3,P_4$ to obtain $\Delta S$, but it turns out that\footnote{To see this, let us take the specific basis $\gamma_i=\begin{pmatrix}
0&\im \sigma_i\\
-\im \sigma_i &0
\end{pmatrix}$ and $\gamma_4=\begin{pmatrix}
0&1\\
1 &0
\end{pmatrix}$. 
Then, the $2$d chirality operator of $T^2$ is given by $\im \gamma_3 \gamma_4=\begin{pmatrix}
-\sigma_3&0\\
0 &\sigma_3
\end{pmatrix}$, and the $2$d massless fermion should take the form $\psi=(0,*,*,0)^t$ to satisfy $\im \gamma_3\gamma_4=1$. This shows explicitly that $\gamma_{3,4}\psi=(*,0,0,*)^t$, which cannot be the zero mode by the spinor structure. }
\begin{equation}
    \overline{\psi}\gamma_{3,4}T^a \psi=0
\end{equation}
when $\psi$ is restricted to $2$d massless fermions satisfying $\im \gamma_3\gamma_4=1$. 
In other words, by emitting gluons in the compactified directions, $P_3$ or $P_4$, the $2$d fermion has to be excited to non-zero KK mode. 
To have the scalar-scalar four-fermion vertex for $2$d massless fermions, we need to consider the loop diagrams, such as 
\begin{align}
	\includegraphics[scale=0.4]{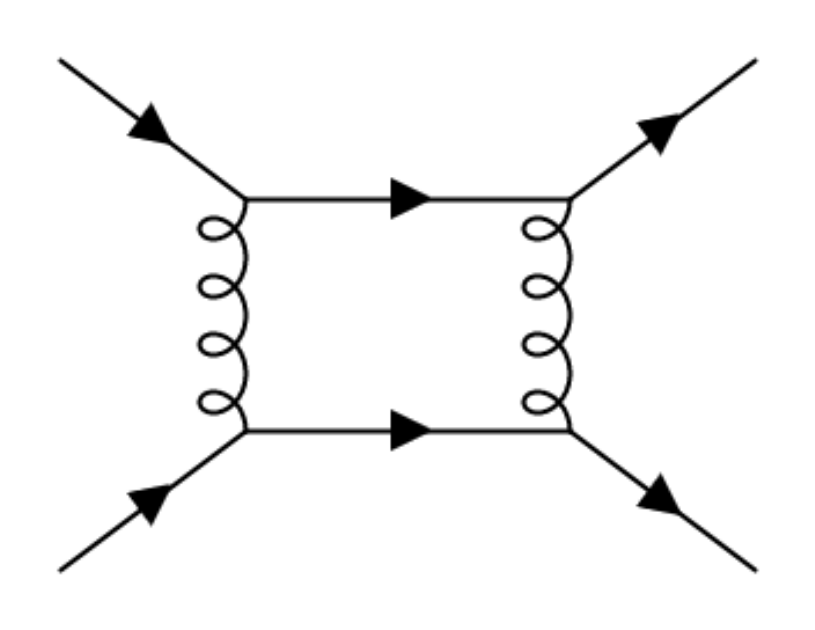},
\end{align}
in which the internal fermion loop runs only over the non-zero KK tower. 
As a result, the coupling $J_{ij}$ in $\Delta S$ is of the order of $g(\mu)^4$, where the renormalization scale should be set as $\mu\sim (NL)^{-1}$. 

As the four-fermion terms~\eqref{eq:4fermi_bosonized} are consistent with the global symmetry, loop diagrams can produce them. 
Instead of performing their explicit calculations, let us simply assume that $\Delta S$ is generated  and that its classical moduli is given by 
\begin{equation}
    \phi_1=\phi_2=\cdots=\phi_{N+1}. 
\end{equation}
Under this assumption, $\Delta S$ gives the marginally relevant perturbation to $S_{\mathrm{tree}}$ according to the renormalization-group analysis of the Berezinskii-Kosterlitz-Thouless (BKT) transition. 
Let us introduce the coordinate of the classical moduli,
\begin{equation}
    \phi\equiv \phi_i,
\end{equation}
and then the effective action of the perturbative gapless mode is given by 
\begin{equation}
    S_{\mathrm{pert.}}=\int_{M}\frac{N+1}{8\pi}|\diff \phi|^2. 
\end{equation}
The zero mode $\phi$ cannot be lifted at any orders of the perturbation theory.

\subsubsection{Reproducing 't~Hooft anomaly with the compact boson}

This is the good moment to take a detour for checking how the anomaly~\eqref{eq:qcdsym_anomaly} is realized in the $2$d effective theory. 
Let $A_{2\rmd}$ and $B_{2\rmd}$ be the $2$d counterpart of background gauge fields defined on $M$, then the background gauge fields on $M\times T^2$ is given by
\begin{equation}
    A=A_{2\rmd}, \, 
    B=B_{2\rmd}+\frac{2\pi}{N}\frac{\diff x_3\wedge \diff x_4}{L^2}. 
\end{equation}
Substituting this expression, the $4$d mixed 't~Hooft anomaly~\eqref{eq:qcdsym_anomaly} becomes 
\begin{equation}
    \bigl(\mathbb{Z}_{2(N+2)}\bigr)_\rmA: \calZ[A_{2\rmd},B_{2\rmd}]\mapsto \rme^{\frac{\im}{(N+2)} \int\left\{(N+1)\diff A_{2\rmd}+NB_{2\rmd}\right\}}\calZ[A_{2\rmd},B_{2\rmd}],
\end{equation}
and this gives the 't~Hooft anomaly for $2$d effective theory. 
The first term comes from the perturbative axial anomaly, and the coefficient $(N+1)$ counts the number of $2$d massless Dirac fermions. 
To reproduce it with compact bosons, we introduce the coupling
\begin{equation}
    \sum_{i=1}^{N+1}\frac{\im}{2\pi}\int_M \phi_i\wedge (\diff A_{2\rmd}+2B_{2\rmd})=
    \frac{\im (N+1)}{2\pi} \int_M\phi\wedge (\diff A_{2\rmd}+2B_{2\rmd}). 
    \label{eq:EFT_bkgd_cpl1}
\end{equation}
This is the standard topological coupling for the vector-like gauge fields in the bosonized description, where the vector-like gauge field strength is made manifestly $1$-form gauge invariant. 
However, this does not reproduce the second term of the anomaly correctly and we need an extra contribution. 
To see where it comes from, we should note that there is the dynamical $\mathbb{Z}_N$ gauge $a'$ with the gas of center-vortex defects: $\diff a'$ can be nonzero and $\int_{V_c} \diff a'=\frac{2\pi}{N}$ for the integration over the vortex core $V_c$. 
This should couple to the $\theta$ parameter in the following standard form, 
\begin{equation}
    \frac{\im \theta}{2\pi}\int_M \diff a', 
\end{equation}
but this does not respect the $U(1)_\rmA$ spurious symmetry, $\theta\to \theta+2(N+2)\alpha$, $\phi \to \phi+2\alpha$, and $\diff a'$ is not  invariant under  $1$-form gauge transformation. 
These requirements fix the form of couplings in a unique manner: 
\begin{equation}
    \frac{\im }{2\pi}\int_{M} (\theta-(N+2)\phi)\wedge (\diff a'+B_{2\rmd}).
    \label{eq:EFT_bkgd_cpl2}
\end{equation}
Combining \eqref{eq:EFT_bkgd_cpl1} and \eqref{eq:EFT_bkgd_cpl2}, the discrete chiral transformation, $\phi\to \phi+\frac{2\pi}{N+2}$, shifts the local counter term as 
\begin{equation}
    \frac{\im}{N+2}\int\left((N+1)(\diff A_{2\rmd}+2B_{2\rmd})-(N+2)B_{2\rmd}\right)=\frac{\im}{N+2}\int\left((N+1)\diff A_{2\rmd}+N B_{2\rmd}\right), 
\end{equation}
and this reproduces the 't~Hooft anomaly. 

\subsubsection{Center vortices and chiral symmetry breaking for QCD(Sym)}

To go beyond the perturbation theory, we should perform the dilute gas approximation of the semiclassical configuration. 
In the $T^2$ compactified setup with the 't~Hooft flux, the BPS bound is given by the center vortex~\cite{Gonzalez-Arroyo:1998hjb, Montero:1999by, Montero:2000pb}, which has the fractional topological charge $Q_{\mathrm{top}}=\frac{1}{N}$ and the nontrivial mutual statistics with Wilson loops.\footnote{
This nontrivial mutual statistics is important to describe confinement using center vortices. 
For numerical studies of center-vortex induced confinement of $4$d gauge theories, see Refs.~\cite{DelDebbio:1996lih,  Faber:1997rp, DelDebbio:1998luz, Langfeld:1998cz, Kovacs:1998xm, Engelhardt:1999fd, deForcrand:1999our}, and for a visual guide, see \cite{Biddle:2019gke}.} 
From the BPS bound, the YM action is given by $S_I/N=\frac{8\pi^2}{g^2 N}$. 
The center-vortex vertex is then given by 
\begin{equation}
    K\rme^{-\frac{S_I}{N}}\rme^{\frac{\im}{N} (\theta-(N+2)\phi)}
    \sim \frac{(\Lambda L)^{3-\frac{4}{3N}}}{L^2} \rme^{\frac{\im}{N} (\theta-(N+2)\phi)}, 
\end{equation}
where $K$ denotes the fluctuation determinant. From the dimensional analysis, $K\sim 1/L^2$. 
This corresponds to the fractionalization of Kobayashi-Maskawa-'t~Hooft vertex~\cite{Kobayashi:1970ji, Kobayashi:1971qz, Maskawa:1974vs, tHooft:1976snw}.\footnote{The six-quark determinant vertex, $\det_{i,j}(\overline{\psi}^i(1+\gamma_5)\psi_j)$, was first introduced in Ref.~\cite{Kobayashi:1970ji} by Kobayashi and Maskawa to phenomenologically explain heavy mass of $\eta'$, and its connection to the ABJ anomaly was speculated in \cite{Maskawa:1974vs}. The concrete mechanism to generate it by instantons was shown by 't~Hooft in Ref.~\cite{tHooft:1976snw}.} 
Accordingly, although the scalar field $\phi$ is introduced as the $2\pi$-periodic field, this expression of the center-vortex vertex does not respect its periodicity. 
We shall see that this problem is resolved by considering the dilute gas of center vortices. 

As the gluon sector is perturbatively gapped due to the 't~Hooft twisted boundary condition, the center vortex has the fixed size,
and the dilute gas approximation is reliable without suffering from infrared divergence. 
Since the 't~Hooft twist is introduced only in the $x_3$-$x_4$ direction, the total topological charge has to be an integer. 
The numbers of center vortices and anti-vortices, $n$ and $\overline{n}$, must satisfy $n-\overline{n}\in N\mathbb{Z}$ for consistency. 
The summation over the dilute vortex gas gives
\begin{align}
    &\sum_{n,\overline{n}\ge 0}\frac{\delta_{n-\overline{n}\in N\mathbb{Z}}}{n! \overline{n}!}
    \left(\int\diff^2 x K\rme^{-\frac{S_I}{N}}\rme^{\frac{\im}{N} (\theta-(N+2)\phi)}\right)^n
    \left(\int\diff^2 x K\rme^{-\frac{S_I}{N}}\rme^{-\frac{\im}{N} (\theta-(N+2)\phi)}\right)^{\overline{n}}\notag\\
    &=\sum_{\ell=0}^{N-1}\exp\left(\int \diff^2 x\, 2K \rme^{-\frac{S_I}{N}}\cos\left(\frac{(N+2)\phi+2\pi \ell-\theta}{N}\right)\right), 
\end{align}
and this full expression satisfy the $2\pi$-periodicity of the field $\phi$. 
However, the local effective potential induced by the center vortex, 
\begin{equation}
    V_{\mathrm{vortex},\ell}(\phi)=-2K\rme^{-\frac{S_I}{N}}\cos\left(\frac{(N+2)\phi+2\pi \ell-\theta}{N}\right),
\end{equation}
has an extra label $\ell\in\mathbb{Z}_N$ and violates the $2\pi$ periodicity for fixed $\ell$. 
A better way to understand its periodicity is given by the following identification,
\begin{align}
    &(\phi, \ell+N)\sim (\phi,\ell),\\
    &(\phi+2\pi,\ell)\sim (\phi, \ell+(N+2))\sim (\phi,\ell+2). 
\end{align}

When $N$ is odd, we can eliminate the label $\ell$ completely by extending the periodicity of $\phi$. That is, the branch label can be absorbed by $\phi$ but the price to pay is to change the periodicity as $\phi\sim \phi+2\pi N$. 
As a result, we can set $\ell=0$, and the ground-states are given by 
\begin{equation}
    \phi=\frac{\theta+2\pi N k}{(N+2)}, 
\end{equation}
with $k\sim k+(N+2)$. Thus, we have $(N+2)$ vacua with $k=0,1,\ldots, (N+2)-1$. 
Accordingly, the chiral condensate of the $k$-th vacuum is given by 
\begin{equation}
\label{chiralcon}
    \langle \overline{\psi}_\rmL\psi_\rmR\rangle_k \sim \Lambda^3 (\Lambda L)^{-\frac{4}{3N}}\rme^{\im \frac{\theta+2\pi N k}{(N+2)}}, 
\end{equation}
and the discrete chiral symmetry is spontaneously broken completely. This result matches 
exactly to the one in QCD(Sym) on $\mathbb R^3 \times S^1$~\cite{Shifman:2008ja}, but interestingly the underlying microscopic mechanisms (center vortices vs. monopole-instantons) are genuinely distinct. This suggest a nontrivial adiabatic continuity between these two reliable semi-classical regimes. 

When $N$ is even, the label $\ell$ cannot be eliminated completely. 
The periodicity of $\phi$ is extended as $\phi\sim \phi+2\pi \frac{N}{2}$, and we still have the discrete label $\ell\in \{0,1\}$. 
The presence of the $\mathbb{Z}_2$ discrete label $\ell$ is the consequence of the $1$-form symmetry, $\mathbb{Z}_2^{[1]}(\subset \mathbb{Z}_N^{[1]})$, and the branch label $\ell$ discriminates the universes\footnote{The $1$-form symmetry in $2$d QFT requires the presence of topological local operators, and we can consider the superselection rule associated with those topological local operators. As the superselection rule in this case is much stronger than the ordinary one for spontaneous symmetry breaking, the corresponding superselction sectors are sometimes called as universes to emphasize it. } in the $2$d effective theory~\cite{Pantev:2005zs, Hellerman:2006zs, Hellerman:2010fv, Tanizaki:2019rbk, Cherman:2019hbq,  Komargodski:2020mxz}. 
That is, the ground states are given by 
\begin{equation}
    (\phi,\ell)=\left(\frac{\theta+2\pi N k}{N+2},0\right), \left(\frac{\theta+2\pi+2\pi N k}{N+2},1\right), 
\end{equation}
with $k\sim k+\frac{N+2}{2}$. There are two universes labelled by $\ell=0,1$, and each universe has $\frac{N+2}{2}$ vacua labelled by $k=0,1,\ldots, \frac{N}{2}$. 
As a result, we get $(N+2)$ ground states, and the chiral symmetry is spontaneously broken completely. 

We note that the fundamental Wilson line is a genuine line operator that transforms under $\mathbb{Z}_2^{[1]}$ when $N$ is even, and the universe label $\ell$ jumps to $\ell+1$ by going across the Wilson line. 
Since two universes have degenerate ground-state energy densities, it obeys the perimeter law, and $\mathbb{Z}_2^{[1]}$ is spontaneously broken. 
This however does not lead extra ground-state degeneracy because of the mixed anomaly between $\mathbb{Z}_2^{[1]}$ and the discrete axial symmetry $(\mathbb{Z}_{2(N+2)})_\rmA$. 
The number of ground states is given by $N+2$ as discussed above.

\subsection{Mass deformation of QCD(Sym) on \texorpdfstring{$M \times T^2$}{M*T2}}

So far, we have considered the massless fermion and obtained the spontaneous chiral symmetry breaking, $\left(\mathbb{Z}_{2(N+2)}\right)_\rmA\xrightarrow{\mathrm{SSB}}\mathbb{Z}_2$, using the semiclassical method. 
Let us add the mass deformation to the Lagrangian, 
\begin{equation}
    \Delta \mathcal{L}_m =  m \left(  (\overline \psi_{\rm L}  \psi_{\rm R}) + (\overline \psi_{\rm R}  \psi_{\rm L})   \right), 
\end{equation} 
and assume that the mass $m$ is sufficiently small. 
The  deformation breaks the 
$\mathbb Z_{2N+4}^{[0]}$ discrete chiral symmetry softly. 
After the bosonization, the mass deformation becomes 
\begin{equation}
    \Delta \mathcal{L}_m\sim -m\frac{(\Lambda L)^{3-\frac{4}{3N}}}{L^3}\cos(\phi). 
\end{equation}
The ground-state energy densities can be obtained by substituting the vacuum configurations for $\phi$ at $m=0$. 

For odd $N$, we then obtain 
\begin{equation}
    E_k(\theta)\sim - m \Lambda^3 (\Lambda L)^{-\frac{4}{3N}}\cos\left(\frac{\theta+2\pi N k}{N+2}\right), 
\end{equation}
with $k=0,1,\ldots, N+1$. At generic values of $\theta$, the system has the unique and gapped ground state. That is, the degeneracy between chirally broken vacua is lifted by the mass deformation. 
Concretely, the ground state is given by $k=0$ when $-\pi<\theta<\pi$. 
At $\theta=\pi$, there are two degenerate vacua, $k=0$ and $Nk=-1$. Here, we note that $Nk=-1$ has the unique solution in $k\in \mathbb{Z}_{N+2}$ as $\gcd(N,N+2)=1$ for odd $N$. 

For even $N$, there are two universes due to $\mathbb{Z}_2^{[1]}$ labelled by $\ell\in \{0,1\}$. 
For the $\ell=0$ universe, the multi-branch ground-state energies are given by 
\begin{equation}
    E^{(\ell=0)}_k(\theta)\sim - m \Lambda^3 (\Lambda L)^{-\frac{4}{3N}} \cos\left(\frac{\theta+2\pi N k}{N+2}\right),
\end{equation}
with $k=0,1,\ldots, \frac{N}{2}$. For the $\ell=1$ universe, they are given by 
\begin{equation}
    E^{(\ell=1)}_{k'}(\theta)\sim - m \Lambda^3 (\Lambda L)^{-\frac{4}{3N}} \cos\left(\frac{\theta+2\pi (N k'+1)}{N+2}\right),
\end{equation}
with $k'=0,1,\ldots, \frac{N}{2}$. 
At generic values of $\theta$, the system again has the unique and gapped ground state. 
As a result, the fundamental Wilson line shows the area law. The string tension $T(\theta)$ of the fundamental Wilson loop is given by 
\begin{equation}
    T(\theta)= L^2\left( \min_{k'}E^{(\ell=1)}_{k'}(\theta)-\min_k E^{(\ell=0)}_{k}(\theta)\right), 
\end{equation}
and it is indeed nonzero for generic values of $\theta$.

\subsection{Semiclassical study of QCD(ASym) on small \texorpdfstring{$\mathbb{R}^2\times T^2$}{R2xT2} with magnetic flux}

In this section, we study the dynamics of QCD(ASym) using the center-vortex effective theory. 
As the discussion is almost parallel to that of QCD(Sym), we explain it very briefly. 

The structure of the vector-like symmetry $G_\rmV$ is completely identical to \eqref{eq:vector_sym} including the form of its background gauge field. 
The ABJ anomaly explicitly breaks $U(1)_\rmA$ to $\left(\mathbb{Z}_{2(N-2)}\right)_\rmA$ since $\tr_{\asym}(T^aT^b)=(N-2)\tr(T^aT^b)$, and thus the factor ``$(N+2)$'' in the analysis of QCD(Sym) should be replaced by ``$(N-2)$'' in the case of QCD(ASym). 
As the dimension of the anti-symmetric representation is $\dim(\asym)=\frac{N(N-1)}{2}$, the factor ``$(N+1)$'' in QCD(Sym) is also replaced by ``$(N-1)$''.

To compute the mixed 't~Hooft anomaly between $G_\rmV$ and the discrete chiral symmetry, let us calculate the index of the Dirac operator under the presence of the $G_\rmV$ background gauge field $(A,B)$:
\begin{align}
    \mathrm{Ind}(\slashed{D})&=\frac{1}{8\pi^2}\int\left(\tr_{\asym}((\widetilde{f}-B)^2)+\dim(\asym)(\diff A+2B)^2\right)\nonumber\\
    &=\frac{1}{8\pi^2}\int\left((N-2)\tr((\widetilde{f}-B)^2)+\frac{N(N-1)}{2}(\diff A+2B)^2\right) \nonumber\\
    &=\frac{N-2}{8\pi^2}\int \tr(\widetilde{f}\wedge \widetilde{f})+\frac{1}{8\pi^2}\int\left(\frac{N(N-1)}{2}(\diff A)^2+2(N-1)\diff A\wedge N B+(NB)^2\right). 
\end{align}
As a result, under the discrete axial transformation, 
\begin{equation}
    \bigl(\mathbb{Z}_{2(N-2)}\bigr)_\rmA:\psi\mapsto \rme^{\frac{2\pi }{2(N-2)}\im\gamma_5}\psi,\, 
    \overline{\psi}\mapsto \overline{\psi}\,\rme^{\frac{2\pi }{2(N-2)}\im\gamma_5}, 
\end{equation}
the partition function of QCD(ASym) is shifted by the local counterterm of the $G_\rmV$ background field:
\begin{align}
    \bigl(\mathbb{Z}_{2(N-2)}\bigr)_\rmA: &\calZ[A,B]\mapsto\rme^{\frac{2\pi \im}{8\pi^2(N-2)} \int\left(\frac{N(N-1)}{2}(\diff A)^2+2(N-1)\diff A\wedge N B+(NB)^2\right)}\calZ[A,B]. 
\end{align}
This is the mixed 't~Hooft anomaly, and the anomaly matching condition can be satisfied by the complete spontaneous breakdown of the discrete axial symmetry, 
\begin{equation}
    \left(\mathbb{Z}_{2(N-2)}\right)_\rmA\xrightarrow{\mathrm{SSB}} \mathbb{Z}_2.
\end{equation}

In order to apply the semiclassical description with center vortices, we would like to perform the $T^2$ compactification to maintain the above 't~Hooft anomaly as much as possible. 
To this end, we again consider the magnetic flux background~\eqref{eq:magnetic_flux}. It effectively introduces the minimal 't~Hooft flux in the gauge sector, and thus the gauge field has the perturbative mass gap because of the Higgsing $SU(N)\to \mathbb{Z}_N$. 
For Dirac fermions, there are perturbative gapless modes following the index theorem on $T^2$ given by \eqref{eq:index_T2}, and it predicts the presence of $(N-1)$ flavors of $2$d Dirac fermions at the classical level. 

By performing the Abelian bosonization, $2$d effective theory at the classical level is given by the theory of $2\pi$-periodic compact scalars, $\phi_i$, with $i=1,\ldots, N-1$. 
The $U(1)_\rmA$ transformation acts as $\phi_i\mapsto \phi_i+2\alpha$, and other axial symmetry in the classical analysis should be eliminated by the four-fermion vertex~\eqref{eq:4fermi_bosonized}, which can be obtained by integrating out massive KK towers. 
We again assume that its classical moduli is given by $\phi\equiv \phi_1=\cdots=\phi_{N-1}$, and then we only have a single compact scalar as a result of the perturbative analysis. 

The center-vortex vertex takes the following form, 
\begin{equation}
    K \rme^{-\frac{S_I}{N}}\rme^{\frac{\im}{N}(\theta-(N-2)\phi)}
    \sim \frac{(\Lambda L)^{3+\frac{4}{3N}}}{L^2}\rme^{\frac{\im}{N}(\theta-(N-2)\phi)},
\end{equation}
and this satisfies the spurious symmetry, $\phi\to \phi+2\alpha$ associated with $\theta\to \theta+2(N-2)\alpha$.  
By performing the dilute gas approximation of center vortices, the effective potential, 
\begin{equation}
    V_{\mathrm{vortex},\ell}(\phi)=-2K\rme^{-\frac{S_I}{N}}\cos\left(\frac{(N-2)\phi+2\pi \ell-\theta}{N}\right),
\end{equation}
is generated with the discrete label, $\ell\in \mathbb{Z}_N$. The periodicity of the field $\phi$ is then understood as 
\begin{align}
    &(\phi,\ell+N)\sim (\phi,\ell), \\
    &(\phi+2\pi,\ell)\sim (\phi,\ell+(N-2))\sim (\phi,\ell-2). 
\end{align}
With this identification, the field $\phi$ is extended to $2\pi N$-periodic field for odd $N$, and the discrete label $\ell$ is completely eliminated. 
For even $N$, the periodicity of $\phi$ is $2\pi \frac{N}{2}$, and $\ell$ takes values in $\mathbb{Z}_2$. In this case, the label $\ell$ discriminates two universes of the $1$-form symmetry, $\mathbb{Z}_2^{[1]}$, in the $2$d effective theory. 
In both cases, we get $(N-2)$ degenerate vacua, which shows the spontaneous breakdown of discrete chiral symmetry. 

\section{Implications on the large-\texorpdfstring{$N$}{N} orbifold/orientifold equivalence}

There is a conjecture called as the large-$N$ orientifold equivalence~\cite{Armoni:2003gp,Armoni:2004ub}, which states that $C$-invariant bosonic sectors of QCD(Sym/ASym) becomes equivalent to those of $\mathcal{N}=1$ $SU(N)$ super Yang-Mills (SYM) theory in the large-$N$ limit. 
This is a striking conjecture, which is interesting and important not only theoretically but also phenomenologically. 
For $N=3$, two-index anti-symmetric representation becomes identical to the fundamental representation, and thus QCD(ASym) for $N=3$ is the ordinary QCD that describes the strong interaction of our world. 
If the conjecture is true and $N=3$ is regarded as a large parameter, the orientifold equivalence may provide a direct path to apply the knowledge of supersymmetric gauge theories to the actual QCD~\cite{Armoni:2003fb, Armoni:2003yv}. 

It is therefore an important question if the large-$N$ orientifold equivalence is true at the nonperturbative level. 
In Ref.~\cite{Unsal:2006pj}, it is uncovered that the unbroken charge-conjugation symmetry is the necessary and sufficient condition for its validity. 
In order to see this, it is useful to think of both $\mathcal{N}=1$ $SU(N)$ SYM and QCD(ASym) as $\mathbb{Z}_2$ projections of $\mathcal{N}=1$ $SO(2N)$ SYM theory. 
To obtain $\mathcal{N}=1$ $SU(N)$ SYM theory from $SO(2N)$ theory, we introduce the symplectic matrix $J=\im \sigma_2\otimes \bm{1}_N\in SO(2N)$ and impose the constraint $A_\mu=J A_\mu J^t$ and $\lambda=J \lambda J^t$ for the gluons and gluinos of $SO(2N)$ theory. 
Then, the gauge group becomes $U(N)$ and $\lambda$ becomes the Weyl fermion in the adjoint representation, so we obtain $\mathcal{N}=1$ $U(N)$ SYM theory. 
To obtain QCD(ASym), we replace the constraint for the gluino as $\lambda=-J \lambda J^t$, then the fermion becomes the Dirac fermion in the anti-symmetric representation of $U(N)$. 
In the large-$N$ limit, the difference between $U(N)$ and $SU(N)$ becomes negligible, and we obtain the correspondence. 
In this way, the large-$N$ orientifold equivalence proposed in Refs.~\cite{Armoni:2003gp, Armoni:2004ub} has been decomposed into two large-$N$ orbifold-type equivalences~\cite{Bershadsky:1998cb, Schmaltz:1998bg, Kachru:1998ys, Strassler:2001fs} between the parent and daughter theories with the same parent theory: It is an example of the daughter-daughter equivalence.

The necessary and sufficient condition for the large-$N$ orbifold equivalence has been derived in Refs.~\cite{Kovtun:2003hr,Kovtun:2004bz,Kovtun:2005kh}, and it shows that the equivalence holds for correlation functions of neutral sectors if and only if the ground state also belongs to the neutral sector. 
Applying it to the present case, the ground states must have unbroken charge-conjugation symmetry, and then the orientifold equivalence is true for observables in the $C$-even bosonic sector. 
Since it is widely accepted that $\mathcal{N}=1$ SYM theory does not break $C$ while it breaks the discrete axial symmetry, the question becomes whether QCD(ASym) maintains $C$-symmetry or not. 
This is a nontrivial question as it involves the strong dynamics of $4$d gauge theories. 

So far, we pay attention to QCD(ASym), but we can perform the same analysis for QCD(Sym) starting from the $\mathcal{N}=1$ $USp(2N)$ SYM theory, and we do not repeat it here. 
Again, the criterion for the large-$N$ orientifold equivalence is given by the unbroken $C$-symmetry for QCD(Sym).

Our analysis on small $\mathbb{R}^2\times T^2$ predicts that $C$ symmetry is unbroken. 
We have also observed the complete spontaneous breakdown of discrete chiral symmetry, which suggests the adiabatic continuity between our semiclassical description and the strong dynamics on $\mathbb{R}^4$. 
This gives positive support for the nonperturbative orientifold equivalence. 

We would like to emphasize the magnetic flux on small $T^2$ plays the pivotal role to achieve the adiabatic continuity~\cite{Tanizaki:2022ngt} (see also Refs.~\cite{vanBaal:1984ra, Yamazaki:2017ulc, Cox:2021vsa}). 
In Ref.~\cite{Unsal:2006pj}, QCD(Sym/ASym) on small $\mathbb{R}^3\times S^1$ has been studied, and it was found that these theories have chiral symmetric vacua with broken $C$ symmetry. 
Since the anomaly matching on $\mathbb{R}^4$ requires the spontaneous breaking of chiral symmetry, this suggests that the naive compactification must have phase transitions and it is not suitable to study the dynamics of strongly coupled regimes. 

In order to achieve the adiabatic continuity on $\mathbb{R}^3\times S^1$, the double-trace deformed gauge theories are more suitable choice~\cite{Unsal:2007vu, Unsal:2007jx, Unsal:2008ch}.
By adding the double-trace deformations to QCD(ASym) on $\mathbb{R}^3\times S^1$, we can find that $C$ becomes unbroken and instead the chiral symmetry is spontaneously broken~\cite{Shifman:2008ja}. 
Therefore, in both $S^1$ and $T^2$ compactifications with adiabatic continuity, we obtained the evidence for unbroken $C$ symmetry, suggesting the validity of nonperturbative large-$N$ orientifold equivalence. 

\section{Summary}
\label{sec:summary}

We performed the semiclassical analysis of QCD(Sym/ASym) on $\mathbb{R}^2\times T^2$ by introducing the magnetic flux in $T^2$. 
QCD(Sym/ASym) have the mixed 't~Hooft anomaly between its vector symmetry and discrete chiral symmetry, and the magnetic flux in $T^2$ allows us to preserve the information of 't~Hooft anomaly. 
As a result, the dynamics of the $2$d effective theory is constrained by the same anomaly of $4$d QCD(Sym/ASym). 

We find that discrete chiral symmetry is spontaneously broken completely thanks to the dilute gas of center vortices, and this satisfies the anomaly matching constraint. 
This leads to the multi-branch structure of the ground states after the tiny mass deformation, and this leads to the same vacuum structure with the pure YM theory. 

Each ground state has unbroken $C$ symmetry, and this gives an important implication for the large-$N$ orientifold equivalence between QCD(Sym/ASym) and $\mathcal{N}=1$ $SU(N)$ SYM theory. 
Since the unbroken $C$ symmetry gives the necessary and sufficient condition for its nonperturbative justification, our result gives a strong positive support for the nonperturbative orientifold equivalence, assuming the adiabatic continuity. 

As we showed in this work, the chiral condensate obtained in the small symmetric $T^2$ regime \eqref{chiralcon} matches exactly the result on $\mathbb R^3 \times S^1_{\rm small}$ in \cite{Shifman:2008ja}. 
At the same time, the microscopic configurations which lead to this result are genuinely distinct, even though they have the same topological charge $Q=1/N$ and action. 
Monopole-instantons and center vortices have different mutual statistics with Wilson loops. It would be very interesting to establish how these two reliable semiclassical regimes are adiabatically connected.

Lastly, let us comment on several future directions. 
One of them is about the domain wall that connects different chiral broken vacua. 
As the phase of chiral condensate can be regarded as an effective $\theta$ angle via the ABJ anomaly, the domain-wall theory can be thought of $3$d QCD(Sym/ASym) with a Chern-Simons term. 
This problem has been studied in Ref.~\cite{Choi:2018tuh}, and they suggest the domain wall connecting $n$-th and $0$-th chiral broken vacua is described by $U(n)_{(N\pm 2)-n,(N\pm 2)-2n}$ Chern-Simons theory, where $n=1,2,\ldots, (n\pm 2)-1$. 
It would be an interesting future work to study if the $T^2$ compactification with magnetic flux can give a nontrivial check of this proposal on the wall. 

Another curious problem is to extend our work of vector-like QCD to some chiral gauge theories~\cite{Bolognesi:2017pek, Bolognesi:2019wfq, Bolognesi:2019fej, Sulejmanpasic:2020zfs}. 
Recently, Ref.~\cite{Anber:2021iip} considers the $4$d chiral gauge theories that contain both two-index symmetric and anti-symmetric Weyl fermions, and the anomaly matching constraint is studied in detail both on spin and non-spin manifolds. 
It would be fascinating if we can check those scenarios (even partially) by the reliable semiclassics with $T^2$ compactification, although the extension might not be straightforward due to the lack of the vector-like $U(1)_\rmB$ symmetry. 

\acknowledgments

This work was completed during the workshop ``Continuous Advances in QCD 2022'' (CAQCD2022), and the authors thank Aleksey Cherman and Misha Shifman for the kind invitation and appreciate their hospitality. 
The work of Y. T. was supported by Japan Society for the Promotion of Science (JSPS) KAKENHI Grant numbers, 22H01218 and 20K22350, and by Center for Gravitational Physics and Quantum Information (CGPQI) at Yukawa Institute for Theoretical Physics.
The work of M. \"{U}. was supported by U.S. Department of Energy, Office of Science, Office of Nuclear Physics under Award Number DE-FG02-03ER41260.

\appendix

\section{Group theoretic factors of two-index representations}

Let $T^a_\rep$ be the generators of $\mathfrak{su}(N)$ Lie algebra with the representation $\rep$. 
We take the orthonormal basis, 
\begin{equation}
    \tr(T^a_\rep T^b_\rep)=T(\rep)\delta^{ab}, 
\end{equation}
where $T(\rep)$ is the trace normalization factor. 
For the defining (or fundamental) representation, $\rep=\mathsf{fd}$, we set 
\begin{equation}
    T(\mathsf{fd})=\frac{1}{2}.
\end{equation}

The quadratic Casimir invariant $C_2(\rep)$ is defined by 
\begin{equation}
    \sum_{a=1}^{N^2-1}T^a_\rep T^a_\rep=C_2(\rep)\, \mathrm{id}_{\rep}. 
\end{equation}
By taking the trace of both sides, $C_2(\rep)$ is related to $T(\rep)$ by the formula,
\begin{equation}
    C_2(\rep)\dim(\rep)=(N^2-1)T(\rep). 
\end{equation}
Let us write down the table of these factors for the fundamental, adjoint, two-index symmetric and antisymmetric representations:
\begin{align}
    \begin{array}{c|ccc}
    \rep & \dim(\rep) & T(\rep)      & C_2(\rep)        \\ \hline
    \fd  & N          & \frac{1}{2}  & \frac{N^2-1}{2N} \\
    \adj & N^2-1      & N            &  N               \\
    \sym &\frac{N(N+1)}{2}&\frac{N+2}{2}&\frac{(N-1)(N+2)}{N} \\
    \asym&\frac{N(N-1)}{2}&\frac{N-2}{2}&\frac{(N+1)(N-2)}{N} 
    \end{array}
\end{align}

The perturbative beta function, $\beta(g)=\frac{\diff g}{\diff \ln \mu}$, is given by 
\begin{equation}
    \beta(g)=-\frac{\beta_0}{(4\pi)^2}g^3-\frac{\beta_1}{(4\pi)^4}g^5+O(g^7).
\end{equation}
The leading coefficient is 
\begin{align}
    \beta_0&=\frac{11}{3}N-\frac{4}{3}T(\rep) N_f, \nonumber\\
    \beta_1&=\frac{34}{3}N^2-\frac{20}{3}N T(\rep)N_f-4C_2(\rep)T(\rep)N_f,
\end{align}
for $SU(N)$ gauge theory with $N_f$ Dirac fermions in the representation $\rep$. 
The normalization of the coefficient is the standard one for non-supersymmetric gauge theories, and then the instanton factor within the $1$-loop renormalization group is given by 
\begin{equation}
    \exp\left(-\frac{8\pi^2}{g^2(\mu)}\right)=\left(\frac{\Lambda}{\mu}\right)^{\beta_0}, 
\end{equation}
where $\Lambda$ is the strong scale. 
For $1$-flavor QCD(Sym/Asym), $\beta_0$ and $\beta_1$ are given by 
\begin{align}
    \beta_0&=3N\mp\frac{4}{3}, \\
    \beta_1&=6N^2\mp \left(\frac{38}{3}N- \frac{8}{N}\right),
\end{align}
and they are identical to those of $\mathcal{N}=1$ super Yang-Mills theory in the large-$N$ limit, $\beta_0=3N$ and $\beta_1=6N^2$.

\section{Dirac zero modes on \texorpdfstring{$T^2$}{T2} with the magnetic flux}
\label{sec:Dirac_zero}

In this Appendix, we confirm the result of the index theorem~\eqref{eq:index_T2} by solving the Dirac zero-mode equation~\eqref{eq:Dirac_zero_mode} explicitly. 
We first define the chirality of small $T^2$ by
\begin{equation}
    (\im \gamma_3\gamma_4)\psi_\pm =\pm \psi_\pm,
\end{equation}
and then the Dirac equation for each chirality becomes 
\begin{equation}
    \left[\partial_3\mp \im \left(\partial_4+\im\frac{4\pi}{N}\frac{x_3}{L^2}\right)\right]\psi_{\pm}=0, 
\end{equation}
or 
\begin{equation}
    (\partial_3\mp \im \partial_4)\left(\rme^{\pm \frac{2\pi}{N}\frac{x_3^2}{L^2}}\psi_{\pm}\right)=0. 
\end{equation}
This shows that 
\begin{equation}
    \psi_\pm(x_3,x_4)=\rme^{\mp \frac{2\pi}{N}\frac{x_3^2}{L^2}}u_\pm ((x_3\mp \im x_4)/L). 
\end{equation}
We can check that there is no zero modes for $\im \gamma_3\gamma_4=-1$ by the following equality, 
\begin{align}
    0&=\int_{T^2} \left|\left(\partial_3\mp \im \left(\partial_4+\im B x_3\right)\right)\psi_{\pm}\right|^2\notag\\
    &=\int_{T^2} \left[|\partial_3 \psi_{\pm}|^2+|(\partial_4+\im B x_3)\psi_{\pm}|^2 \mp B |\psi_{\pm}|^2\right]. 
\end{align}
Here, we set $B=\frac{4\pi}{NL^2}$ for simplicity of notation. 
In the first equality, we use the equation for Dirac zero modes. The second equality can be obtained using integration by parts. For $\im\gamma_3\gamma_4=-1$, all the three terms are positive semi-definite, and it requires that $\psi_-=0$.
We therefore only consider the case $\im \gamma_3\gamma_4=1$ in the following discussion.

Introducing the dimensionless coordinate $z=(x_3-\im x_4)/L$, we can rewrite the boundary condition as  
\begin{align}
	&u_+(z+1)
	=\sym(S^\dagger)\rme^{\frac{2\pi}{N}+\frac{4\pi}{N}z}u_+(z),\\
	&u_+(z-\im)=\sym(C^\dagger)u_+(z). 
\end{align}
Writing color components explicitly, we find that 
\begin{align}
    &u_{+,ab}(z+1)=\rme^{\frac{2\pi}{N}+\frac{4\pi}{N}z}u_{+,a-1\,b-1}(z), \label{eq:bc_x3dir}\\
    &u_{+,ab}(z-\im)=\rme^{-\frac{2\pi\im}{N}((a-1)+(b-1))}u_{+,ab}(z). \label{eq:bc_x4dir} 
\end{align}
We can perform the Fourier expansion to satisfy \eqref{eq:bc_x4dir}, 
\begin{equation}
    u_{+,ab}(z)=\rme^{\frac{2\pi}{N}((a-1)+(b-1))z}\sum_{n}c_{n,ab}\rme^{2\pi n z}. 
\end{equation}
We then use the boundary condition~\eqref{eq:bc_x3dir} for the $x_3$ direction. When $1<a,b\le N$, the condition becomes
\begin{equation}
    \rme^{\frac{2\pi}{N}(a-1+b-1)(z+1)}\sum_n c_{n,ab}\rme^{2\pi n (z+1)}
    =\rme^{\frac{2\pi}{N}+\frac{4\pi}{N}z}\rme^{\frac{2\pi}{N}(a-2+b-2)z}\sum_n c_{n,a-1\, b-1}\rme^{2\pi n z},
\end{equation}
and we have 
\begin{equation}
    c_{n,ab}=\rme^{\frac{2\pi}{N}(1-(a-1)-(b-1))-2\pi n}c_{n,a-1\, b-1}. 
\end{equation}
When $a=1$ but $b>1$, the condition becomes 
\begin{equation}
    \rme^{\frac{2\pi}{N}(0+b-1)(z+1)}\sum_n c_{n,1\, b}\rme^{2\pi n(z+1)}
    =\rme^{\frac{2\pi}{N}+\frac{4\pi}{N}z}\rme^{\frac{2\pi}{N}(N-1+b-2)z}\sum_n c_{n,N\, b-1}\rme^{2\pi n z}, 
\end{equation}
and we have 
\begin{equation}
    c_{n,1\, b}=\rme^{\frac{2\pi}{N}(1-(b-1))-2\pi n}c_{n-1, N\, b-1}.
\end{equation}
Similarly, when $a>1$ and $b=1$, 
\begin{equation}
    c_{n,a\, 1}=\rme^{\frac{2\pi}{N}(1-(a-1))-2\pi n}c_{n-1, a-1\, N}. 
\end{equation}
When $a=b=1$, we get 
\begin{equation}
    c_{n,11}=\rme^{\frac{2\pi}{N}-2\pi n}c_{n-2, NN}. 
\end{equation}
This gives $c_{n,a\,b}\propto \exp\left(-\frac{2\pi N}{4}n^2-\pi((a-1)+(b-1))n\right)$, and we can find that the solutions can be expressed with theta functions.\footnote{The Fourier series expansion of Jacobi's theta function is given by $\theta_3(z)\equiv \theta_3(z|\tau)=\sum_{n\in \mathbb{Z}}\rme^{\im \pi \tau n^2+2\im n z}=1+2\sum_{n\ge 1}\rme^{\im \pi \tau n^2}\cos(2 n z)$ with $\mathrm{Im}(\tau)>0$. Importantly, it satisfies the quasi-periodicity, $\theta_3(z+(m+n\tau)\pi)=\rme^{-\im \pi \tau n^2-2\im n z}\theta_3(z)$, and it is the key property to solve the quasi-periodicity of $u_+$ coming out of \eqref{eq:bc_x3dir}, $u_+(z+N)=\rme^{2\pi N+4\pi z}u_+(z)$, combined with \eqref{eq:bc_x4dir}. 
We note that the theta functions are holomorphic, i.e. do not have poles or singularities, and thus they give normalizable zero modes.
}

Let us count the number of zero modes. By the shift matrix, $a-b \bmod N$ is preserved, so the diagonal directions are related in mod $N$. That means, when we consider $k$-diagonal from the top-left to the bottom-right (i.e. $c_{n,i\, i+k}$ wit $i:1\to N-k$), we jump to the $-(N-k)$-diagonal elements (i.e. $c_{n+1,N-k+j\, j}$ with $j:1\to k$), and we come back to the $k$-diagonal elements, $c_{n+2,1\, 1+k}$. 
As a result, $c_{n, ab}$ is related to $c_{n\pm 2,ab}$, which means that we can have $2$ zero modes, corresponding to even and odd $n$: For example, the $1$-$1$ component is given by 
\begin{equation}
    u_{+,11}(z)=a\sum_{n\in 2\mathbb{Z}}\rme^{-\frac{2\pi N}{4}n^2}\rme^{2\pi n z}+b \sum_{n\in 2\mathbb{Z}+1}\rme^{-\frac{2\pi N}{4}n^2}\rme^{2\pi n z},
\end{equation}
where $a,b\in \mathbb{C}$. 
For the main diagonal, we thus have $2$ zero modes, but for off-diagonals, the symmetry condition imposes the restriction, and thus we only have $1$ zero mode for each. In total, we get $2+(N-1)=N+1$ zero modes as indicated by the index theorem~\eqref{eq:index_T2}. 

The discussion is completely parallel also for the case of the two-index anti-symmetric representation. The crucial difference appears only in the last step and it comes from the absence of the diagonal elements. As a result, we get $0+(N-1)=N-1$ zero modes.

\bibliographystyle{utphys}
\bibliography{./QFT.bib,./refs.bib}

\providecommand{\href}[2]{#2}\begingroup\raggedright\begin{thebibliography}{10}

\bibitem{Tanizaki:2022ngt}
Y.~Tanizaki and M.~\"Unsal, ``{Center vortex and confinement in Yang-Mills
  theory and QCD with anomaly-preserving compactifications},''
  \href{http://dx.doi.org/10.1093/ptep/ptac042}{{\em PTEP} {\bfseries 2022}
  (2022) 04A108}, \href{http://arxiv.org/abs/2201.06166}{{\ttfamily
  arXiv:2201.06166 [hep-th]}}.

\bibitem{tHooft:1979rtg}
G.~'t~Hooft, ``{A Property of Electric and Magnetic Flux in Nonabelian Gauge
  Theories},''
\href{http://dx.doi.org/10.1016/0550-3213(79)90595-9}{{\em Nucl. Phys.}
  {\bfseries B153} (1979) 141--160}.

\bibitem{tHooft:1977nqb}
G.~'t~Hooft, ``On the phase transition towards permanent quark confinement,''
  \href{http://dx.doi.org/10.1016/0550-3213(78)90153-0}{{\em Nucl.Phys.B}
  {\bfseries 138} (1978) 1--25}.

\bibitem{Cornwall:1979hz}
J.~M. Cornwall, ``{Quark Confinement and Vortices in Massive Gauge Invariant
  QCD},'' \href{http://dx.doi.org/10.1016/0550-3213(79)90111-1}{{\em Nucl.
  Phys. B} {\bfseries 157} (1979) 392--412}.

\bibitem{Nielsen:1979xu}
H.~B. Nielsen and P.~Olesen, ``{A Quantum Liquid Model for the QCD Vacuum:
  Gauge and Rotational Invariance of Domained and Quantized Homogeneous Color
  Fields},'' \href{http://dx.doi.org/10.1016/0550-3213(79)90065-8}{{\em Nucl.
  Phys. B} {\bfseries 160} (1979) 380--396}.

\bibitem{Ambjorn:1980ms}
J.~Ambjorn and P.~Olesen, ``{A Color Magnetic Vortex Condensate in QCD},''
  \href{http://dx.doi.org/10.1016/0550-3213(80)90150-9}{{\em Nucl. Phys. B}
  {\bfseries 170} (1980) 265--282}.

\bibitem{Gonzalez-Arroyo:1998hjb}
A.~Gonzalez-Arroyo and A.~Montero, ``{Selfdual vortex - like configurations in
  SU(2) Yang-Mills theory},''
  \href{http://dx.doi.org/10.1016/S0370-2693(98)01229-5}{{\em Phys. Lett. B}
  {\bfseries 442} (1998) 273--278},
  \href{http://arxiv.org/abs/hep-th/9809037}{{\ttfamily arXiv:hep-th/9809037}}.

\bibitem{Montero:1999by}
A.~Montero, ``{Study of SU(3) vortex - like configurations with a new maximal
  center gauge fixing method},''
  \href{http://dx.doi.org/10.1016/S0370-2693(99)01113-2}{{\em Phys. Lett. B}
  {\bfseries 467} (1999) 106--111},
  \href{http://arxiv.org/abs/hep-lat/9906010}{{\ttfamily
  arXiv:hep-lat/9906010}}.

\bibitem{Montero:2000pb}
A.~Montero, ``{Vortex configurations in the large N limit},''
  \href{http://dx.doi.org/10.1016/S0370-2693(00)00572-4}{{\em Phys. Lett. B}
  {\bfseries 483} (2000) 309--314},
  \href{http://arxiv.org/abs/hep-lat/0004002}{{\ttfamily
  arXiv:hep-lat/0004002}}.

\bibitem{tHooft:1973alw}
G.~'t~Hooft, ``{A Planar Diagram Theory for Strong Interactions},''
\href{http://dx.doi.org/10.1016/0550-3213(74)90154-0}{{\em Nucl. Phys.}
  {\bfseries B72} (1974) 461}.

\bibitem{Armoni:2003gp}
A.~Armoni, M.~Shifman, and G.~Veneziano, ``{Exact results in non-supersymmetric
  large N orientifold field theories},''
  \href{http://dx.doi.org/10.1016/S0550-3213(03)00538-8}{{\em Nucl. Phys.}
  {\bfseries B667} (2003) 170--182},
\href{http://arxiv.org/abs/hep-th/0302163}{{\ttfamily arXiv:hep-th/0302163
  [hep-th]}}.

\bibitem{Armoni:2004ub}
A.~Armoni, M.~Shifman, and G.~Veneziano, ``{Refining the proof of planar
  equivalence},'' \href{http://dx.doi.org/10.1103/PhysRevD.71.045015}{{\em
  Phys. Rev.} {\bfseries D71} (2005) 045015},
\href{http://arxiv.org/abs/hep-th/0412203}{{\ttfamily arXiv:hep-th/0412203
  [hep-th]}}.

\bibitem{Unsal:2006pj}
M.~Unsal and L.~G. Yaffe, ``{(In)validity of large N orientifold
  equivalence},'' \href{http://dx.doi.org/10.1103/PhysRevD.74.105019}{{\em
  Phys. Rev. D} {\bfseries 74} (2006) 105019},
  \href{http://arxiv.org/abs/hep-th/0608180}{{\ttfamily arXiv:hep-th/0608180}}.

\bibitem{Shifman:2008ja}
M.~Shifman and M.~Unsal, ``{QCD-like Theories on R(3) x S(1): A Smooth Journey
  from Small to Large r(S(1)) with Double-Trace Deformations},''
  \href{http://dx.doi.org/10.1103/PhysRevD.78.065004}{{\em Phys. Rev.}
  {\bfseries D78} (2008) 065004},
\href{http://arxiv.org/abs/0802.1232}{{\ttfamily arXiv:0802.1232 [hep-th]}}.

\bibitem{Gaiotto:2017yup}
D.~Gaiotto, A.~Kapustin, Z.~Komargodski, and N.~Seiberg, ``{Theta, Time
  Reversal, and Temperature},''
  \href{http://dx.doi.org/10.1007/JHEP05(2017)091}{{\em JHEP} {\bfseries 05}
  (2017) 091},
\href{http://arxiv.org/abs/1703.00501}{{\ttfamily arXiv:1703.00501 [hep-th]}}.

\bibitem{Tanizaki:2017bam}
Y.~Tanizaki and Y.~Kikuchi, ``{Vacuum structure of bifundamental gauge theories
  at finite topological angles},''
  \href{http://dx.doi.org/10.1007/JHEP06(2017)102}{{\em JHEP} {\bfseries 06}
  (2017) 102},
\href{http://arxiv.org/abs/1705.01949}{{\ttfamily arXiv:1705.01949 [hep-th]}}.

\bibitem{Komargodski:2017dmc}
Z.~Komargodski, A.~Sharon, R.~Thorngren, and X.~Zhou, ``{Comments on Abelian
  Higgs Models and Persistent Order},''
  \href{http://dx.doi.org/10.21468/SciPostPhys.6.1.003}{{\em SciPost Phys.}
  {\bfseries 6} no.~1, (2019) 003},
\href{http://arxiv.org/abs/1705.04786}{{\ttfamily arXiv:1705.04786 [hep-th]}}.

\bibitem{Komargodski:2017smk}
Z.~Komargodski, T.~Sulejmanpasic, and M.~Unsal, ``{Walls, anomalies, and
  deconfinement in quantum antiferromagnets},''
  \href{http://dx.doi.org/10.1103/PhysRevB.97.054418}{{\em Phys. Rev.}
  {\bfseries B97} no.~5, (2018) 054418},
\href{http://arxiv.org/abs/1706.05731}{{\ttfamily arXiv:1706.05731
  [cond-mat.str-el]}}.

\bibitem{Shimizu:2017asf}
H.~Shimizu and K.~Yonekura, ``{Anomaly constraints on deconfinement and chiral
  phase transition},'' \href{http://dx.doi.org/10.1103/PhysRevD.97.105011}{{\em
  Phys. Rev.} {\bfseries D97} no.~10, (2018) 105011},
\href{http://arxiv.org/abs/1706.06104}{{\ttfamily arXiv:1706.06104 [hep-th]}}.

\bibitem{Gaiotto:2017tne}
D.~Gaiotto, Z.~Komargodski, and N.~Seiberg, ``{Time-reversal breaking in
  QCD$_{4}$, walls, and dualities in 2 + 1 dimensions},''
  \href{http://dx.doi.org/10.1007/JHEP01(2018)110}{{\em JHEP} {\bfseries 01}
  (2018) 110},
\href{http://arxiv.org/abs/1708.06806}{{\ttfamily arXiv:1708.06806 [hep-th]}}.

\bibitem{Kapustin:2014gua}
A.~Kapustin and N.~Seiberg, ``{Coupling a QFT to a TQFT and Duality},''
  \href{http://dx.doi.org/10.1007/JHEP04(2014)001}{{\em JHEP} {\bfseries 04}
  (2014) 001},
\href{http://arxiv.org/abs/1401.0740}{{\ttfamily arXiv:1401.0740 [hep-th]}}.

\bibitem{Tanizaki:2018wtg}
Y.~Tanizaki, ``{Anomaly constraint on massless QCD and the role of Skyrmions in
  chiral symmetry breaking},''
  \href{http://dx.doi.org/10.1007/JHEP08(2018)171}{{\em JHEP} {\bfseries 08}
  (2018) 171},
\href{http://arxiv.org/abs/1807.07666}{{\ttfamily arXiv:1807.07666 [hep-th]}}.

\bibitem{Anber:2019nze}
M.~M. Anber and E.~Poppitz, ``{On the baryon-color-flavor (BCF) anomaly in
  vector-like theories},''
  \href{http://dx.doi.org/10.1007/JHEP11(2019)063}{{\em JHEP} {\bfseries 11}
  (2019) 063}, \href{http://arxiv.org/abs/1909.09027}{{\ttfamily
  arXiv:1909.09027 [hep-th]}}.

\bibitem{Anber:2021lzb}
M.~M. Anber, ``{Condensates and anomaly cascade in vector-like theories},''
  \href{http://dx.doi.org/10.1007/JHEP03(2021)191}{{\em JHEP} {\bfseries 03}
  (2021) 191}, \href{http://arxiv.org/abs/2101.04132}{{\ttfamily
  arXiv:2101.04132 [hep-th]}}.

\bibitem{Tanizaki:2017qhf}
Y.~Tanizaki, T.~Misumi, and N.~Sakai, ``{Circle compactification and 't Hooft
  anomaly},'' \href{http://dx.doi.org/10.1007/JHEP12(2017)056}{{\em JHEP}
  {\bfseries 12} (2017) 056},
\href{http://arxiv.org/abs/1710.08923}{{\ttfamily arXiv:1710.08923 [hep-th]}}.

\bibitem{Tanizaki:2017mtm}
Y.~Tanizaki, Y.~Kikuchi, T.~Misumi, and N.~Sakai, ``{Anomaly matching for phase
  diagram of massless $\mathbb{Z}_N$-QCD},''
  \href{http://dx.doi.org/10.1103/PhysRevD.97.054012}{{\em Phys. Rev.}
  {\bfseries D97} (2018) 054012},
\href{http://arxiv.org/abs/1711.10487}{{\ttfamily arXiv:1711.10487 [hep-th]}}.

\bibitem{Yamazaki:2017dra}
M.~Yamazaki, ``{Relating 't Hooft Anomalies of 4d Pure Yang-Mills and 2d
  $\mathbb{CP}^{N-1}$ Model},''
  \href{http://dx.doi.org/10.1007/JHEP10(2018)172}{{\em JHEP} {\bfseries 10}
  (2018) 172},
\href{http://arxiv.org/abs/1711.04360}{{\ttfamily arXiv:1711.04360 [hep-th]}}.

\bibitem{DelDebbio:1996lih}
L.~Del~Debbio, M.~Faber, J.~Greensite, and S.~Olejnik, ``{Center dominance and
  Z(2) vortices in SU(2) lattice gauge theory},''
  \href{http://dx.doi.org/10.1103/PhysRevD.55.2298}{{\em Phys. Rev. D}
  {\bfseries 55} (1997) 2298--2306},
  \href{http://arxiv.org/abs/hep-lat/9610005}{{\ttfamily
  arXiv:hep-lat/9610005}}.

\bibitem{Faber:1997rp}
M.~Faber, J.~Greensite, and S.~Olejnik, ``{Casimir scaling from center
  vortices: Towards an understanding of the adjoint string tension},''
  \href{http://dx.doi.org/10.1103/PhysRevD.57.2603}{{\em Phys. Rev. D}
  {\bfseries 57} (1998) 2603--2609},
  \href{http://arxiv.org/abs/hep-lat/9710039}{{\ttfamily
  arXiv:hep-lat/9710039}}.

\bibitem{DelDebbio:1998luz}
L.~Del~Debbio, M.~Faber, J.~Giedt, J.~Greensite, and S.~Olejnik, ``{Detection
  of center vortices in the lattice Yang-Mills vacuum},''
  \href{http://dx.doi.org/10.1103/PhysRevD.58.094501}{{\em Phys. Rev. D}
  {\bfseries 58} (1998) 094501},
  \href{http://arxiv.org/abs/hep-lat/9801027}{{\ttfamily
  arXiv:hep-lat/9801027}}.

\bibitem{Langfeld:1998cz}
K.~Langfeld, O.~Tennert, M.~Engelhardt, and H.~Reinhardt, ``{Center vortices of
  Yang-Mills theory at finite temperatures},''
  \href{http://dx.doi.org/10.1016/S0370-2693(99)00252-X}{{\em Phys. Lett. B}
  {\bfseries 452} (1999) 301},
  \href{http://arxiv.org/abs/hep-lat/9805002}{{\ttfamily
  arXiv:hep-lat/9805002}}.

\bibitem{Kovacs:1998xm}
T.~G. Kovacs and E.~T. Tomboulis, ``{Vortices and confinement at weak
  coupling},'' \href{http://dx.doi.org/10.1103/PhysRevD.57.4054}{{\em Phys.
  Rev. D} {\bfseries 57} (1998) 4054--4062},
  \href{http://arxiv.org/abs/hep-lat/9711009}{{\ttfamily
  arXiv:hep-lat/9711009}}.

\bibitem{Engelhardt:1999fd}
M.~Engelhardt, K.~Langfeld, H.~Reinhardt, and O.~Tennert, ``{Deconfinement in
  SU(2) Yang-Mills theory as a center vortex percolation transition},''
  \href{http://dx.doi.org/10.1103/PhysRevD.61.054504}{{\em Phys. Rev. D}
  {\bfseries 61} (2000) 054504},
  \href{http://arxiv.org/abs/hep-lat/9904004}{{\ttfamily
  arXiv:hep-lat/9904004}}.

\bibitem{deForcrand:1999our}
P.~de~Forcrand and M.~D'Elia, ``{On the relevance of center vortices to QCD},''
  \href{http://dx.doi.org/10.1103/PhysRevLett.82.4582}{{\em Phys. Rev. Lett.}
  {\bfseries 82} (1999) 4582--4585},
  \href{http://arxiv.org/abs/hep-lat/9901020}{{\ttfamily
  arXiv:hep-lat/9901020}}.

\bibitem{Biddle:2019gke}
J.~C. Biddle, W.~Kamleh, and D.~B. Leinweber, ``{Visualization of center vortex
  structure},'' \href{http://dx.doi.org/10.1103/PhysRevD.102.034504}{{\em Phys.
  Rev. D} {\bfseries 102} no.~3, (2020) 034504},
  \href{http://arxiv.org/abs/1912.09531}{{\ttfamily arXiv:1912.09531
  [hep-lat]}}.

\bibitem{Kobayashi:1970ji}
M.~Kobayashi and T.~Maskawa, ``{Chiral symmetry and eta-x mixing},''
\href{http://dx.doi.org/10.1143/PTP.44.1422}{{\em Prog. Theor. Phys.}
  {\bfseries 44} (1970) 1422--1424}.

\bibitem{Kobayashi:1971qz}
M.~Kobayashi, H.~Kondo, and T.~Maskawa, ``{Symmetry breaking of the chiral u(3)
  x u(3) and the quark model},''
  \href{http://dx.doi.org/10.1143/PTP.45.1955}{{\em Prog. Theor. Phys.}
  {\bfseries 45} (1971) 1955--1959}.

\bibitem{Maskawa:1974vs}
T.~Maskawa and H.~Nakajima, ``{Spontaneous Symmetry Breaking in Vector-Gluon
  Model},'' \href{http://dx.doi.org/10.1143/PTP.52.1326}{{\em Prog. Theor.
  Phys.} {\bfseries 52} (1974) 1326--1354}.

\bibitem{tHooft:1976snw}
G.~'t~Hooft, ``{Computation of the Quantum Effects Due to a Four-Dimensional
  Pseudoparticle},'' \href{http://dx.doi.org/10.1103/PhysRevD.18.2199.3,
  10.1103/PhysRevD.14.3432}{{\em Phys. Rev.} {\bfseries D14} (1976)
  3432--3450}.
[,70(1976)].

\bibitem{Pantev:2005zs}
T.~Pantev and E.~Sharpe, ``{GLSM's for Gerbes (and other toric stacks)},''
  \href{http://dx.doi.org/10.4310/ATMP.2006.v10.n1.a4}{{\em Adv. Theor. Math.
  Phys.} {\bfseries 10} no.~1, (2006) 77--121},
\href{http://arxiv.org/abs/hep-th/0502053}{{\ttfamily arXiv:hep-th/0502053
  [hep-th]}}.

\bibitem{Hellerman:2006zs}
S.~Hellerman, A.~Henriques, T.~Pantev, E.~Sharpe, and M.~Ando, ``{Cluster
  decomposition, T-duality, and gerby CFT's},''
  \href{http://dx.doi.org/10.4310/ATMP.2007.v11.n5.a2}{{\em Adv. Theor. Math.
  Phys.} {\bfseries 11} no.~5, (2007) 751--818},
  \href{http://arxiv.org/abs/hep-th/0606034}{{\ttfamily arXiv:hep-th/0606034}}.

\bibitem{Hellerman:2010fv}
S.~Hellerman and E.~Sharpe, ``{Sums over topological sectors and quantization
  of Fayet-Iliopoulos parameters},''
  \href{http://dx.doi.org/10.4310/ATMP.2011.v15.n4.a7}{{\em Adv. Theor. Math.
  Phys.} {\bfseries 15} (2011) 1141--1199},
\href{http://arxiv.org/abs/1012.5999}{{\ttfamily arXiv:1012.5999 [hep-th]}}.

\bibitem{Tanizaki:2019rbk}
Y.~Tanizaki and M.~Unsal, ``{Modified instanton sum in QCD and
  higher-groups},'' \href{http://dx.doi.org/10.1007/JHEP03(2020)123}{{\em JHEP}
  {\bfseries 03} (2020) 123},
\href{http://arxiv.org/abs/1912.01033}{{\ttfamily arXiv:1912.01033 [hep-th]}}.

\bibitem{Cherman:2019hbq}
A.~Cherman, T.~Jacobson, Y.~Tanizaki, and M.~\"Unsal, ``{Anomalies, a mod 2
  index, and dynamics of 2d adjoint QCD},''
  \href{http://dx.doi.org/10.21468/SciPostPhys.8.5.072}{{\em SciPost Phys.}
  {\bfseries 8} (2020) 072}, \href{http://arxiv.org/abs/1908.09858}{{\ttfamily
  arXiv:1908.09858 [hep-th]}}.

\bibitem{Komargodski:2020mxz}
Z.~Komargodski, K.~Ohmori, K.~Roumpedakis, and S.~Seifnashri, ``{Symmetries and
  strings of adjoint QCD$_{2}$},''
  \href{http://dx.doi.org/10.1007/JHEP03(2021)103}{{\em JHEP} {\bfseries 03}
  (2021) 103}, \href{http://arxiv.org/abs/2008.07567}{{\ttfamily
  arXiv:2008.07567 [hep-th]}}.

\bibitem{Armoni:2003fb}
A.~Armoni, M.~Shifman, and G.~Veneziano, ``{SUSY relics in one flavor QCD from
  a new 1/N expansion},''
  \href{http://dx.doi.org/10.1103/PhysRevLett.91.191601}{{\em Phys. Rev. Lett.}
  {\bfseries 91} (2003) 191601},
\href{http://arxiv.org/abs/hep-th/0307097}{{\ttfamily arXiv:hep-th/0307097
  [hep-th]}}.

\bibitem{Armoni:2003yv}
A.~Armoni, M.~Shifman, and G.~Veneziano, ``{QCD quark condensate from SUSY and
  the orientifold large N expansion},''
  \href{http://dx.doi.org/10.1016/j.physletb.2003.10.094}{{\em Phys. Lett.}
  {\bfseries B579} (2004) 384--390},
\href{http://arxiv.org/abs/hep-th/0309013}{{\ttfamily arXiv:hep-th/0309013
  [hep-th]}}.

\bibitem{Bershadsky:1998cb}
M.~Bershadsky and A.~Johansen, ``{Large N limit of orbifold field theories},''
  \href{http://dx.doi.org/10.1016/S0550-3213(98)00526-4}{{\em Nucl. Phys.}
  {\bfseries B536} (1998) 141--148},
\href{http://arxiv.org/abs/hep-th/9803249}{{\ttfamily arXiv:hep-th/9803249
  [hep-th]}}.

\bibitem{Schmaltz:1998bg}
M.~Schmaltz, ``{Duality of nonsupersymmetric large N gauge theories},''
  \href{http://dx.doi.org/10.1103/PhysRevD.59.105018}{{\em Phys. Rev.}
  {\bfseries D59} (1999) 105018},
\href{http://arxiv.org/abs/hep-th/9805218}{{\ttfamily arXiv:hep-th/9805218
  [hep-th]}}.

\bibitem{Kachru:1998ys}
S.~Kachru and E.~Silverstein, ``{4-D conformal theories and strings on
  orbifolds},'' \href{http://dx.doi.org/10.1103/PhysRevLett.80.4855}{{\em Phys.
  Rev. Lett.} {\bfseries 80} (1998) 4855--4858},
\href{http://arxiv.org/abs/hep-th/9802183}{{\ttfamily arXiv:hep-th/9802183
  [hep-th]}}.

\bibitem{Strassler:2001fs}
M.~J. Strassler, ``{On methods for extracting exact nonperturbative results in
  nonsupersymmetric gauge theories},''
\href{http://arxiv.org/abs/hep-th/0104032}{{\ttfamily arXiv:hep-th/0104032
  [hep-th]}}.

\bibitem{Kovtun:2003hr}
P.~Kovtun, M.~Unsal, and L.~G. Yaffe, ``{Nonperturbative equivalences among
  large N(c) gauge theories with adjoint and bifundamental matter fields},''
  \href{http://dx.doi.org/10.1088/1126-6708/2003/12/034}{{\em JHEP} {\bfseries
  12} (2003) 034},
\href{http://arxiv.org/abs/hep-th/0311098}{{\ttfamily arXiv:hep-th/0311098
  [hep-th]}}.

\bibitem{Kovtun:2004bz}
P.~Kovtun, M.~Unsal, and L.~G. Yaffe, ``{Necessary and sufficient conditions
  for non-perturbative equivalences of large N(c) orbifold gauge theories},''
  \href{http://dx.doi.org/10.1088/1126-6708/2005/07/008}{{\em JHEP} {\bfseries
  07} (2005) 008},
\href{http://arxiv.org/abs/hep-th/0411177}{{\ttfamily arXiv:hep-th/0411177
  [hep-th]}}.

\bibitem{Kovtun:2005kh}
P.~Kovtun, M.~Unsal, and L.~G. Yaffe, ``{Can large N(c) equivalence between
  supersymmetric Yang-Mills theory and its orbifold projections be valid?},''
  \href{http://dx.doi.org/10.1103/PhysRevD.72.105006}{{\em Phys. Rev.}
  {\bfseries D72} (2005) 105006},
\href{http://arxiv.org/abs/hep-th/0505075}{{\ttfamily arXiv:hep-th/0505075
  [hep-th]}}.

\bibitem{vanBaal:1984ra}
P.~van Baal, ``{Twisted Boundary Conditions: A Nonperturbative Probe for Pure
  Nonabelian Gauge Theories},'' {Ph.D. thesis}, {Rijksuniversiteit Utrecht},
  1984.

\bibitem{Yamazaki:2017ulc}
M.~Yamazaki and K.~Yonekura, ``{From 4d Yang-Mills to 2d $\mathbb{CP}^{N-1}$
  model: IR problem and confinement at weak coupling},''
  \href{http://dx.doi.org/10.1007/JHEP07(2017)088}{{\em JHEP} {\bfseries 07}
  (2017) 088},
\href{http://arxiv.org/abs/1704.05852}{{\ttfamily arXiv:1704.05852 [hep-th]}}.

\bibitem{Cox:2021vsa}
A.~A. Cox, E.~Poppitz, and F.~D. Wandler, ``{The mixed 0-form/1-form anomaly in
  Hilbert space: pouring the new wine into old bottles},''
  \href{http://dx.doi.org/10.1007/JHEP10(2021)069}{{\em JHEP} {\bfseries 10}
  (2021) 069}, \href{http://arxiv.org/abs/2106.11442}{{\ttfamily
  arXiv:2106.11442 [hep-th]}}.

\bibitem{Unsal:2007vu}
M.~Unsal, ``{Abelian duality, confinement, and chiral symmetry breaking in
  QCD(adj)},'' \href{http://dx.doi.org/10.1103/PhysRevLett.100.032005}{{\em
  Phys. Rev. Lett.} {\bfseries 100} (2008) 032005},
\href{http://arxiv.org/abs/0708.1772}{{\ttfamily arXiv:0708.1772 [hep-th]}}.

\bibitem{Unsal:2007jx}
M.~Unsal, ``{Magnetic bion condensation: A New mechanism of confinement and
  mass gap in four dimensions},''
  \href{http://dx.doi.org/10.1103/PhysRevD.80.065001}{{\em Phys. Rev.}
  {\bfseries D80} (2009) 065001},
\href{http://arxiv.org/abs/0709.3269}{{\ttfamily arXiv:0709.3269 [hep-th]}}.

\bibitem{Unsal:2008ch}
M.~Unsal and L.~G. Yaffe, ``{Center-stabilized Yang-Mills theory: Confinement
  and large N volume independence},''
  \href{http://dx.doi.org/10.1103/PhysRevD.78.065035}{{\em Phys. Rev.}
  {\bfseries D78} (2008) 065035},
\href{http://arxiv.org/abs/0803.0344}{{\ttfamily arXiv:0803.0344 [hep-th]}}.

\bibitem{Choi:2018tuh}
C.~Choi, D.~Delmastro, J.~Gomis, and Z.~Komargodski, ``{Dynamics of QCD$_{3}$
  with Rank-Two Quarks And Duality},''
  \href{http://dx.doi.org/10.1007/JHEP03(2020)078}{{\em JHEP} {\bfseries 03}
  (2020) 078}, \href{http://arxiv.org/abs/1810.07720}{{\ttfamily
  arXiv:1810.07720 [hep-th]}}.

\bibitem{Bolognesi:2017pek}
S.~Bolognesi, K.~Konishi, and M.~Shifman, ``{Patterns of symmetry breaking in
  chiral QCD},'' \href{http://dx.doi.org/10.1103/PhysRevD.97.094007}{{\em Phys.
  Rev.} {\bfseries D97} no.~9, (2018) 094007},
\href{http://arxiv.org/abs/1712.04814}{{\ttfamily arXiv:1712.04814 [hep-th]}}.

\bibitem{Bolognesi:2019wfq}
S.~Bolognesi and K.~Konishi, ``{Dynamics and symmetries in chiral $SU(N)$ gauge
  theories},''
\href{http://arxiv.org/abs/1906.01485}{{\ttfamily arXiv:1906.01485 [hep-th]}}.

\bibitem{Bolognesi:2019fej}
S.~Bolognesi, K.~Konishi, and A.~Luzio, ``{Gauging 1-form center symmetries in
  simple $SU(N)$ gauge theories},''
\href{http://arxiv.org/abs/1909.06598}{{\ttfamily arXiv:1909.06598 [hep-th]}}.

\bibitem{Sulejmanpasic:2020zfs}
T.~Sulejmanpasic, Y.~Tanizaki, and M.~\"{U}nsal, ``{Universality between
  vector-like and chiral quiver gauge theories: Anomalies and domain walls},''
  \href{http://dx.doi.org/10.1007/JHEP06(2020)173}{{\em JHEP} {\bfseries 06}
  (2020) 173}, \href{http://arxiv.org/abs/2004.10328}{{\ttfamily
  arXiv:2004.10328 [hep-th]}}.

\bibitem{Anber:2021iip}
M.~M. Anber, S.~Hong, and M.~Son, ``{New anomalies, TQFTs, and confinement in
  bosonic chiral gauge theories},''
  \href{http://dx.doi.org/10.1007/JHEP02(2022)062}{{\em JHEP} {\bfseries 02}
  (2022) 062}, \href{http://arxiv.org/abs/2109.03245}{{\ttfamily
  arXiv:2109.03245 [hep-th]}}.

\end{thebibliography}\endgroup
\end{document}